Review

Roman Y. Shopa* and Kamil Dulski

# Multi-photon time-of-flight MLEM application for the positronium imaging in J-PET

**Abstract:** We develop a positronium imaging method for the Jagiellonian PET (J-PET) scanners based on the time-of-flight maximum likelihood expectation maximisation (TOF MLEM). The system matrix elements are calculated on-the-fly for the coincidences comprising two annihilation and one de-excitation photons that originate from the ortho-positronium (o-Ps) decay. Using the Geant4 library, a Monte Carlo simulation was conducted for four cylindrical $^{22}$Na sources of $\beta^+$ decay with diverse o-Ps mean lifetimes, placed symmetrically inside the two JPET prototypes. The estimated time differences between the annihilation and the positron emission were aggregated into histograms (one per voxel), updated by the weights of the activities reconstructed by TOF MLEM. The simulations were restricted to include only the o-Ps decays into back-to-back photons, allowing a linear fitting model to be employed for the estimation of the mean lifetime from each histogram built in the log scale. To suppress the noise, the exclusion of voxels with activity below 2% – 10% of the peak was studied. The estimated o-Ps mean lifetimes were consistent with the simulation and distributed quasi-uniformly at high MLEM iterations. The proposed positronium imaging technique can be further upgraded to include various correction factors, as well as be modified according to realistic o-Ps decay models.

## Introduction

The main application of the positron emission tomography (PET) in nuclear medicine is the diagnostics and the monitoring treatment of the pathological lesions that can be tracked by the uptake of radio-tracers specific for a particular anomaly or a disease, such as cancer or Alzheimer's [1, 2]. In general, a PET image, reconstructed from the detected 511-keV photon pairs that originate from a positron-electron ($e^+e^-$) annihilation, targets the metabolic processes in tissues, when an abnormally high levels may point at a neoplastic lesion. However, PET might as well be employed to study structural impurities via positron annihilation lifetime spectroscopy (PALS) – a technique based on the decay of positronium (Ps). It is a meta-stable hydrogen-like compound of $e^+$ and $e^-$ that can exist in two states, depending on spin – para-positronium (p-Ps) and ortho-positronium (o-Ps) [3, 4]. The self-annihilation lifetimes of p-Ps and o-Ps in vacuum are 125 ps and 142 ns, respectively. The latter value is much shorter inside tissues due to a limited size of the void o-Ps is located at and an interaction with the neighbouring molecular material. Two processes can lead to swift $e^+e^-$ annihilation: picking up an electron with a spin opposite to a positron or via conversion o-Ps → p-Ps (e.g. by interacting with an oxygen). That reduces the effective o-Ps lifetimes to 1.8 ns in water and to 4 ns inside skin cells [3, 5]. Positronium imaging is a technique that reflects such regularities inside the studied object based on Ps mean lifetime [6].

PALS studies conducted so far reported promising results. Hypoxic tumours, resistant to chemotherapy and radiation treatments, have the O$_2$ concentration different from the control tissues and can be identified from o-Ps lifetimes [7–9]. More recently, a correlation was found between the mean lifetimes obtained by PALS and via histopathological analysis of diseased tissues [10–14]. However, no decent PET studies

___________

***Corresponding author: Roman Y. Shopa,** Department of Complex Systems, National Centre for Nuclear Research, Otwock-Świerk, Poland, E-mail: roman.shopa@ncbj.gov.pl. https://orcid.org/0000-0002-1089-5050
**Kamil Dulski,** Faculty of Physics, Astronomy and Applied Computer Science, Jagiellonian University, Kraków, Poland; Center for Theranostics, Jagiellonian University, Kraków, Poland; and INFN, Laboratori Nazionali di Frascati, Frascati, Italy. E-mail: kamil.dulski@gmail.com

have been conducted until recently, although up to 40% of the $e^+e^-$ annihilation during a PET scan originate from Ps decay, and o-Ps is formed in about 75% of these cases [4].

The first experimental results for positronium imaging were obtained by the 3-layer "Big barrel" Jagiellonian PET (J-PET) prototype [15]. J-PET is a novel technology that uses plastic scintillator strips to detect $e^+e^-$ annihilation photons via Compton scattering, with time-of-flight (TOF) information available [1, 15–20]. A generic algorithm – standardised uptake value (SUV) – was used to reconstruct the expected positions of $e^+e^-$ annihilation and build Ps lifetime (PL) spectra, similarly to the studies made on simulated data (see [21–23]). However, the time resolution of the Big barrel (assessed as 460 ps for a three-photon coincidence), its axial uncertainty of detection ($\sigma_Z \approx 15$ mm, see [24]) and sensitivity lower than in conventional scanners, do not favour such an approach. Despite the existing J-PET machinery is capable of effective positronium imaging in clinics, a proper imaging technique has not been developed so far [25]. Recent studies propose algorithms based on maximum likelihood (ML), yet they utilise an oversimplified decay model for PL spectra [26–28].

In this work, we redefine an iterative TOF ML expectation maximization (TOF MLEM) image reconstruction method for three-photon coincidence events originating from the ortho-positronium (o-Ps) decays with the pick-off or conversion to p-Ps. The algorithm is based on the one developed for the conventional PET scans that account for the resolution model in J-PET [29]. Using the simulated data, we replicate the real experiment with four point-like sources placed inside the Big barrel scanner and in a newer, 24-module J-PET prototype. The statistical nature of the upgraded TOF MLEM allows for building smoother histograms used to determine o-Ps mean lifetimes, which reconstructed distribution exhibits much less noise and blur compared to SUV approach.

# Methodology

### Geant4 simulation setup

By utilising dedicated software[1] based on the Geant4 libraries [30], we conducted a Monte Carlo simulation, replicating the experiment reported in the recent work [15]. A $^{22}$Na isotope was used as a source of β$^+$ decay via the following reaction: $^{22}$Na → $^{22}$Ne$^*$ + $e^+$ + $\nu$ → $^{22}$Ne + $\gamma_{1274}$ + $e^+$ + $\nu$. Here, $\gamma_{1274}$ denotes the deexcitation (prompt) photon of the energy 1274 keV.

We simulated two existing J-PET prototypes with the following geometries (Figure 1, a):

– "Big barrel": 3-layer cylindrical scanner (radii 425 mm, 467.5 mm, 575 mm) with 192 sparsely arranged detector strips of the size 7 mm × 19 mm × 500 mm [19].

– Modular: a tomograph of a radius 381.86 mm constituted by 24 modules each comprising 13 strips with dimensions 6 mm × 24 mm × 456 mm [1].

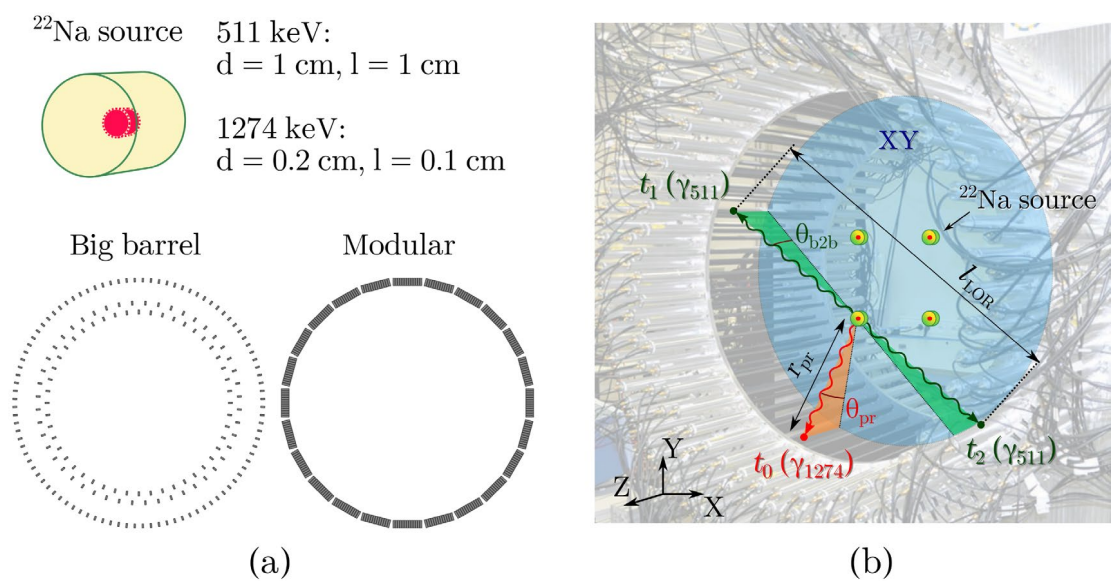

**Figure 1:** Source dimensions and transverse views on the simulated J-PET scanners (a); schematic depiction of an emission from one of the four simulated $^{22}$Na sources and its detection inside the Big barrel prototype (b).

---

[1] https://github.com/JPETTomography/J-PET-geant4

Four small cylindrical sources with uniform emission probabilities and total activity 1.1 MBq were defined, symmetrically located inside the scanners at $x$ = ±81 mm, $y$ = ±81 mm, z=0 mm (Figure 1 – larger cylinder emulates positron range). Two sources had a slightly longer o-Ps lifetime. The simulations were restricted to the o-Ps decays that produced pairs of 511-keV back-to-back annihilation photons as the result of either a pick-off process or a conversion of o-Ps to p-Ps via a neighbouring nucleus and its following self-decay [31]. To acquire exactly 3 hits (scatterings in scintillators) per coincidence – a 511-keV pair and 1274-keV prompt, – the selection criteria were applied, using a 20-ns time windows and the same time-over-thresholds and geometrical relations between the positions of hits as in the previous study [15]. Out of 21×10⁶ events simulated in each scanner, only about 25 000 matched the criteria.

An exemplary coincidence with two back-to-back 511-keV photons produced as a result of $e^+e^-$ annihilation is shown in Figure 1, b. The line-of-response (LOR) which these photons propagate along has an obliqueness angle $\theta_{\text{b2b}}$, different from the obliqueness $\theta_{\text{pr}}$ of the 1274-keV prompt. The difference between the emission times of the back-to-back pair t$_{b2b}$ and the deexcitation photon t$_{pr}$ is used to build a PL spectrum for each voxel. Using the notation in Figure 1, b, this difference is calculated as follows:

$$\Delta t = t_{\text{b2b}} - t_{\text{pr}}, \quad (1)$$

$$t_{\text{pr}} = t_0 - r_{\text{pr}}/c_0,$$

$$t_{\text{b2b}} = 0.5 \cdot (t_1 + t_2 - l_{\text{LOR}}/c_0),$$

where $c_0$ is the speed of light, usually set as in vacuum.

## TOF MLEM for three-photon PET

To estimate the unknown PET activity distribution in some voxel $\lambda_j$ after n-th iteration, the following list-mode MLEM update formula is used:

$$\lambda_j^{(n+1)} = \frac{\lambda_j^{(n)}}{\sum_{i \in I} m_{ij}} \sum_{\epsilon \in \mathcal{E}} \frac{m_{i_\epsilon,j}}{\sum_{j' \in J} m_{i_\epsilon,j'} \lambda_{j'}^{(n)}}, \quad (2)$$

where each J-PET system matrix element $m_{ij}$ reflects the probability that a 3-photon coincidence event originated from the j-th voxel was registered by the i-th combination of detection elements (bins). The summation in list-mode is done over the measured events $\epsilon \in \mathcal{E}$ [32]. A multi-photon $m_{ij}$ can be decomposed into simpler bins and account for TOF and axial blur using two 1D Gaussian kernels – $H_{\text{TOF}}(\cdot)$ and $H_z(\cdot)$, respectively [29]. One system matrix element will then be transformed as follows:

$$m_{ij} \to m_{k_\epsilon,j}^{(\text{b2b})} \cdot m_{l_\epsilon,j}^{(\text{pr})} \cdot H_{\text{TOF}}(\Delta l_{\epsilon j}) \cdot H_z(\Delta z_{\epsilon j}), (3)$$

where $i_\epsilon = k_\epsilon \cap l_\epsilon$, $m_{k_\epsilon,j}^{(\text{b2b})}$ and $m_{l_\epsilon,j}^{(\text{pr})}$ are the detection probabilities for the back-to-back photon pair and the deexcitation (prompt) photon, respectively, $\Delta l_{\epsilon j}$ is the distance from the j-th voxel to the annihilation point projected to LOR and $\Delta z_{\epsilon j}$ – the axial offset of the same voxel measured from LOR along Z-axis.

Considering the decomposition (3), the sensitivity correction factor $\sum_i m_{ij}$ in the denominator of (2) can be estimated using two separate Monte Carlo simulations for $m_{kj}^{(\text{b2b})}$ and $m_{lj}^{(\text{pr})}$ (see [29]). However, it is difficult and unpractical to precalculate $m_{ij}$ elements for all possible three-photon bins $i \in I$, so an event-by-event approach is a reasonable alternative.

## On-the-fly system matrix modelling

The complex J-PET geometry produces significant detector blur that affects its system matrix. Since the traditional pre-computation is expensive, we propose to precalculate $m_{i_\epsilon,j}$ for each event inside a limited volume around the annihilation point. Providing the smooth character of the J-PET strips, this can be done in 2D, projecting each LOR for a 511-keV back-to-back pair onto XY-plane and adjusting the attenuation factors according to the obliqueness angles $\theta_{\text{b2b}}$ and $\theta_{\text{pr}}$ (see Figure 1, b). Next, we allocate small rectangular subregions around the annihilation point aligned along LOR (Figure 2). Their number is restricted by the TOF kernel (±3.5$\sigma_{\text{TOF}}$ for $\theta_{\text{b2b}}$=0, where $\sigma_{\text{TOF}}$ is a standard deviation (SD) of TOF resolution rescaled into units of length) and detector blur (finite cross-sections of the J-PET strips).

In order to estimate the detection probability for every subregion, a set of Siddon projectors is defined, each rotated by a small angle $\delta\varphi$ [33]. The raytracing lines of these projectors must cross the J-PET strips that registered the event: one recording the prompt photon and two – the back-to-back pair (the line should not exceed the "blurred" LOR – see Figure 2). The estimated probabilities are then integrated over $\delta\varphi$ and multiplied for all three. The eventually built low-resolution 2D "heatmap" of each event is stored in memory along with the obliqueness $\theta_{\text{b2b}}$ and reloaded upon each MLEM iter-

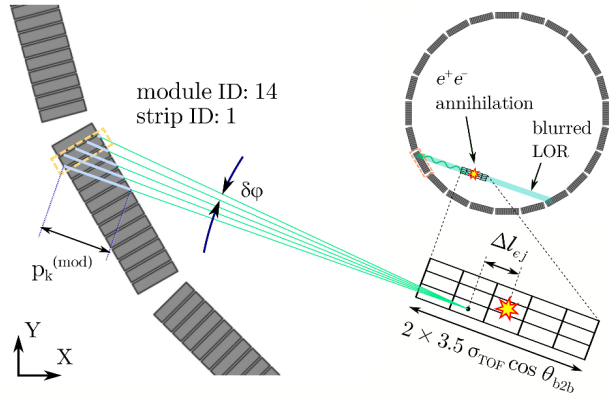

**Figure 2:** Precalculation of the system matrix elements using allocated transverse subregions and raytracing inside the modular J-PET scanner. Each probability of a k-th projector (in green) considers attenuation across the path $p_k^{(mod)}$.

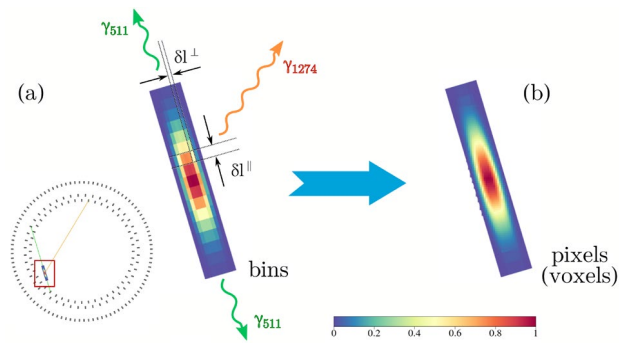

**Figure 3:** Exemplary precomputed transverse 2D "heatmap" of detection probabilities (a) and the resulting image updated by bilinear interpolation (b) for one event inside the Big barrel J-PET prototype.

-ation (Figure 3, a). To update voxel intensities $\lambda_j^{(n)}$, a bilinear interpolation is employed as shown in Figure 3, b for the "heatmap" projected back to LOR using $\theta_{b2b}$. Finally, the axial kernel $H_Z(\Delta z_{\epsilon j})$ is imposed to update the neighbouring voxels.

## Estimation of o-Ps lifetime from the PL spectrum

During the final TOF MLEM iteration, the time differences $\Delta t$ from (1) are calculated for each voxel inside the region around the annihilation point (see Figure 3, b) and added to the corresponding histograms (one per voxel) as the intensity weights taken from the reconstructed PET image.

At the next step, each PL spectrum is fitted by a model constituted by the linear combination of the exponentially modified Gaussian profiles, each representing a different decay mechanism [15]:

$$\hat{f}(\Delta t) = \sum_{i=1}^{4} I_i \exp\left(\frac{\Delta t}{\tau_i}\right) * H_{\text{Gauss}}(\Delta t, \mu, \sigma) + \text{Background} \quad (4)$$

where $\tau_i$ and $I_i$ are the mean lifetime and the amplitude of the i-th component, respectively, $\mu$ is a mean and $\sigma$ is a SD of the Gaussian function $H_{\text{Gauss}}(\cdot)$.

As mentioned above, we restricted the simulated model to o-Ps decays only, with the longest mean lifetime $\tau_i$. That allowed us to employ a linear fitting model, providing the spectrum $\hat{f}(\Delta t)$ for each voxel is in log-scale. The only issue here was to restrict the time region for $\Delta t$ to cover a linear part (Figure 4). After the inferential analysis of the data, it was set as [22.5 ns, 397.5 ns].

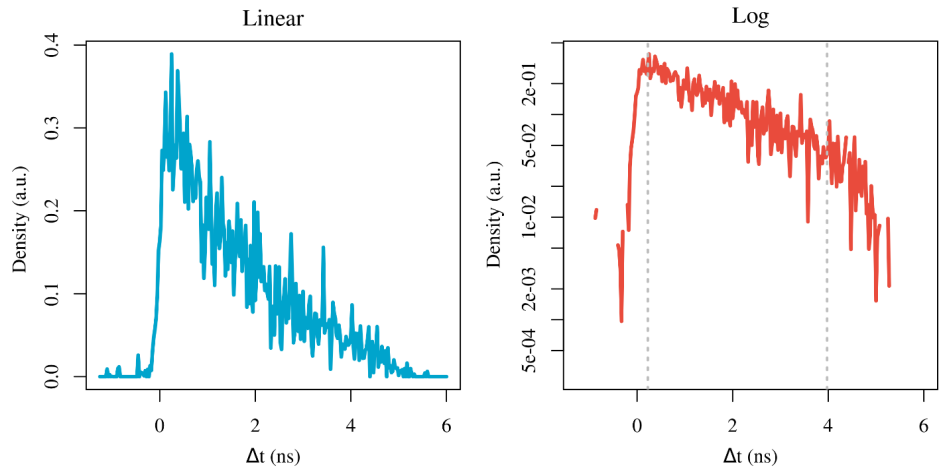

**Figure 4:** Exemplary PL spectrum built for a single-component o-Ps decay model employed for the Geant4 simulation, shown in linear and log scale. The dotted grey lines mark the region where a linear fit is applied.

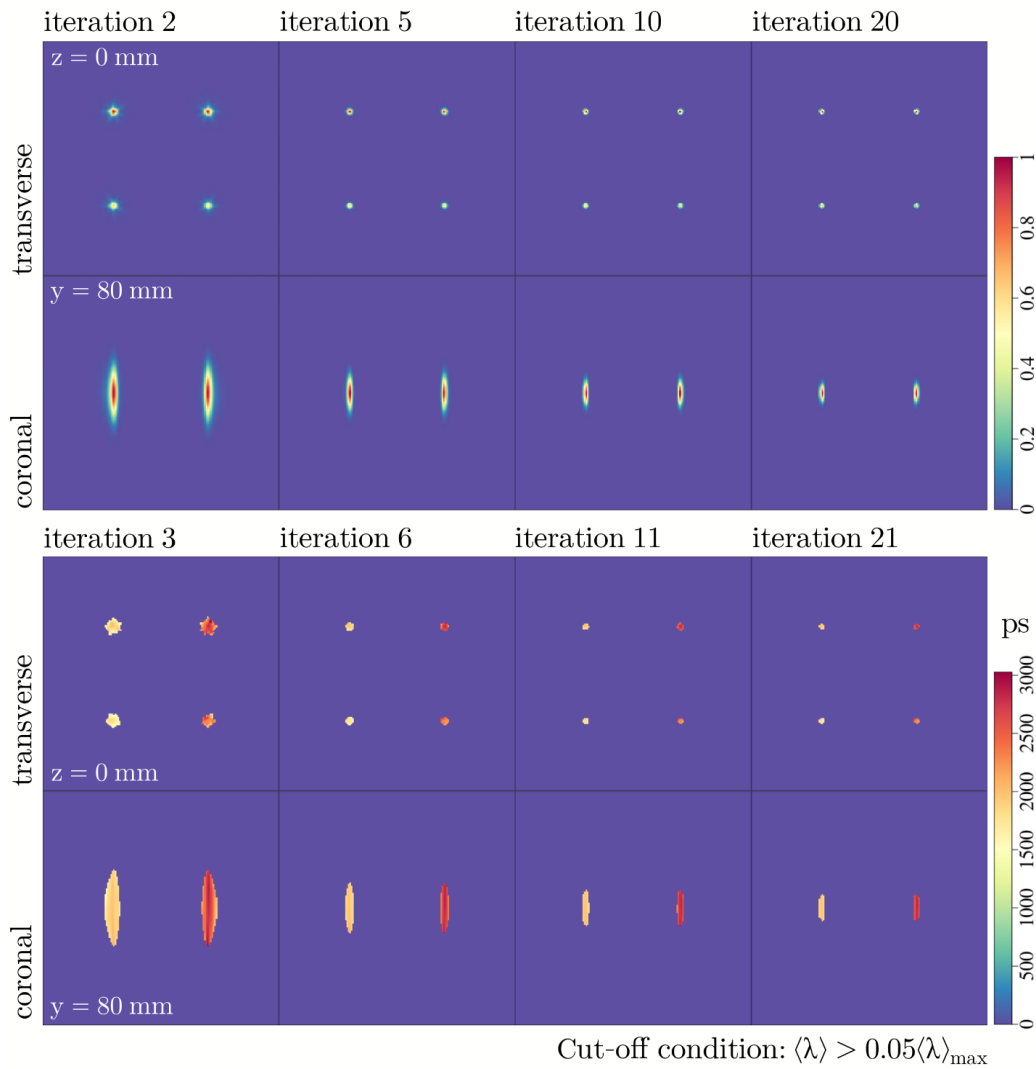

**Figure 5:** Reconstructed by TOF MLEM activity concentrations $\langle\lambda\rangle^{(n)}$ (iterations $n$ = 2, 5, 10, 20) of the simulated sources in the Big barrel J-PET (top) and the corresponding o-Ps mean lifetime images $\langle\tau\rangle^{(n+1)}$ built at the next iteration for the voxels that exceed 5% of $\langle\lambda\rangle_{max}$ (bottom). The cross-sections cover the 400 mm × 400 mm areas at the centre of the field-of-view.

## Results and discussion

We set the voxel size for all images as 2.5 mm × 2.5 mm × 2.5 mm. The on-the-fly system matrix elements for both scanners were calculated using the following parameters: $\delta\varphi$=0.0025, $\delta l^{\perp}$=2.3 mm, $\delta l^{\parallel}$=7.5 mm (see Figure 3, a) $\sigma_{TOF}$=25 mm and $\sigma_Z$=15 mm. That reflects the time resolution ~400 ps.

The Figure 5 shows the cross-sections of the four sources in the Big barrel J-PET, reconstructed by the TOF MLEM (shown on top as voxel activities $\langle\lambda\rangle$) and after having assessed the o-Ps mean lifetimes $\langle\tau\rangle$ from the PL histograms built at the next iteration (bottom). The latter spectra were built only for the voxels with activities higher than 5% of the peak intensity $\langle\lambda\rangle_{max}$. The cut-off was aimed at filtering out the data, insufficient to acquire a plausible histogram.

As we see, the TOF MLEM application improves the spatial resolution for both PET and o-Ps lifetime images, with reduced noise and relatively smooth $\langle\tau\rangle$ distribution, as expected for the simulated data. Compared to the SUV approach utilised in earlier works [15, 22, 23], the application of TOF MLEM is a definite step forward in multi-photon imaging techniques.

The optimal cut-off threshold levels depend on the iteration, as well as on the temporal and spatial resolution. We conducted an exploratory study adjusting filtering conditions for PET images $\langle\lambda\rangle$. One example is depicted in Figure 6 – for the o-Ps mean lifetimes $\langle\tau\rangle^{(6)}$ ($n$=6 denotes the iteration number) estimated for the simulated data in a modular J-PET. Note the difference in the locations of the sources with larger $\langle\tau\rangle$ compared to the Big barrel study.

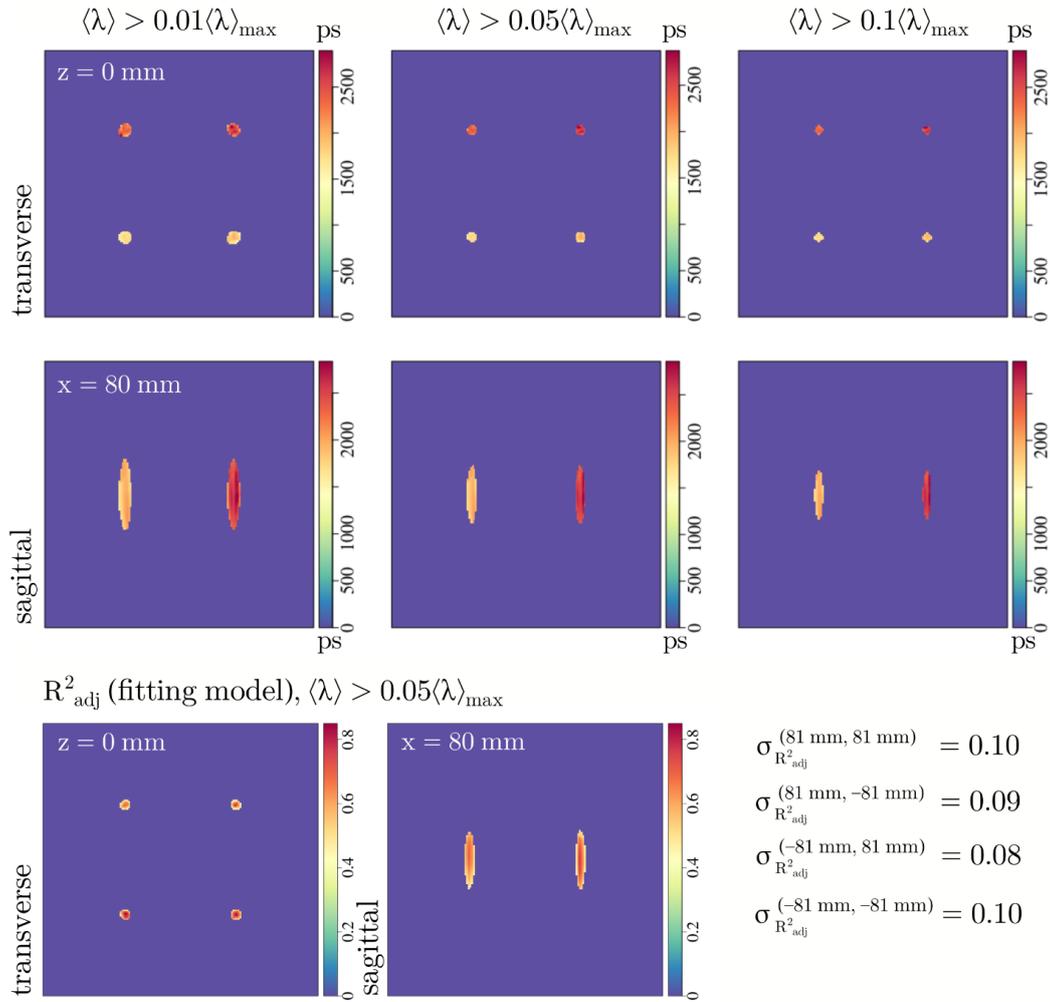

**Figure 6:** The role of the cut-off threshold applied to the reconstructed activities $\langle\lambda\rangle$ for building the o-Ps mean lifetime images $\langle\tau\rangle^{(6)}$ (at the 6-th MLEM iteration), for the four sources simulated in the modular J-PET.

| Scanner | $(x_{src}, y_{src})$, mm $z_{src} = 0$ mm | $\lambda_{max}^{(GT)}$, a.u. | $\langle\lambda\rangle_{max}^{(7)}$, a.u. | $\tau^{(GT)}$, ns | $\langle\tau\rangle_{max}^{(8)}$, ns | $\langle\tau\rangle_{mean}^{(8)}$, ns | $\sigma_{\langle\tau\rangle^{(8)}}$, ns |
|---|---|---|---|---|---|---|---|
| Big barrel | (81, 81) | 1.0 | 1.0 | 2.65 | 3.09 | 2.59 | 0.18 |
|  | (81, −81) | 0.59 | 0.55 | 2.58 | 2.75 | 2.30 | 0.16 |
|  | (−81, 81) | 0.90 | 0.98 | 1.95 | 2.06 | 1.85 | 0.10 |
|  | (−81, −81) | 0.62 | 0.61 | 1.87 | 1.86 | 1.67 | 0.11 |
| Modular | (81, 81) | 1.0 | 0.90 | 2.65 | 2.92 | 2.51 | 0.16 |
|  | (81, −81) | 1.0 | 1.0 | 1.92 | 2.12 | 1.85 | 0.12 |
|  | (−81, 81) | 1.0 | 0.87 | 2.55 | 2.67 | 2.29 | 0.12 |
|  | (−81, −81) | 1.0 | 0.99 | 1.85 | 1.82 | 1.71 | 0.07 |

**Table 1:** Intensities and o-Ps mean lifetimes, estimated for the four reconstructed sources in J-PET scanners at the 7-th and 8-th iteration, respectively. The cut-off threshold was set at 2% of $\langle\lambda\rangle_{max}$.

If the cut-off is set too low, the number of voxels that constitute the $\langle\tau\rangle$ image is higher, and it takes excessively more time to process the PL spectra with sparse and noisy data. As seen for the lowest threshold of 1% shown in Figure 6 (top left), the outcomes are noisier and the shapes of the sources are badly preserved. On the other hand, setting the cut-off too high might result in a loss of important data.

To refine the criteria, the performance of the fitting model can be used, e.g. by tracking the coefficient of determination $R^2$. The exemplary case for the adjusted $R^2_{\text{adj}}$ is shown on the bottom row in Figure 6. We observe a significant variance, confirmed by local SDs $\sigma^{(x_{\text{src}},y_{\text{src}})}_{R^2_{\text{adj}}}$ ($x_{\text{src}}$ and $y_{\text{src}}$ denote source location), which may mean that even a 5%-threshold is too low. Arguably, this issue requires further research for bigger phantoms to make reasonable conclusions.

Table 1 presents the comparison between the reconstructed peak activities $\langle\lambda\rangle^{(7)}_{\max}$ of the sources and the simulated ones $\lambda^{(\text{GT})}_{\max}$ (GT denotes ground truth), as well as between the corresponding mean lifetimes $\langle\tau\rangle^{(8)}_{\max}$ and $\tau^{(\text{GT})}$ (fixed across the volume of each source). The values were calculated at the 7-th iteration for PET and the 8-th – for PL imaging. As expected, the images $\langle\lambda\rangle$ were smoother, hence their normalised peak intensities were in good agreement with what was defined in the Geant4 setup. In comparison, the peak lifetimes $\langle\tau\rangle_{\max}$ did not exhibit such a consistency with the simulation, so we added the averages $\langle\tau\rangle_{\text{mean}}$ and the corresponding SDs $\sigma_{\langle\tau\rangle^{(8)}}$ (last two columns), estimated for the non-zero GT voxels around each simulated source, i.e. under condition $\lambda^{(\text{GT})}>0$. One could argue, though, that, despite $\langle\tau\rangle^{(8)}_{\text{mean}}$ were in better agreement with $\tau^{(\text{GT})}$, such comparison might not be practical for large phantoms.

## Conclusions

We developed a hybrid reconstruction technique for the positronium imaging in J-PET based on the list-mode TOF MLEM that utilises a three-photon system response model that accounts for TOF, detector blur and axial smearing of hits. The first testing was performed on the simulated data for the small cylindrical sources inside the J-PET prototypes, using a simplified o-Ps decay model. The need for filtering out the voxels with low activity concentrations obtained by the TOF MLEM is demonstrated, which allows capitalising on the resolution recovery. The reconstructed images of the mean o-Ps lifetimes exhibited low noise and were systematically consistent with the Geant4 setup. The original algorithm can be further upgraded to incorporate complex decay models, include sensitivity and attenuation corrections, as well as be utilised for positronium imaging in clinics.

## Research funding

We acknowledge the support by the Foundation for Polish Science through TEAM POIR.04.04.00-00-4204/17, the National Science Centre of Poland (through grants No. 2021/41/N/ST2/03950, 2021/42/A/ST2/00423, 2021/43/B/ST2/02150), the Ministry of Education and Science through grant No. SPUB/SP/490528/2021, the Jagiellonian University via project CRP/0641.221.2020, as well as the SciMat and qLife Priority Research Areas budget under the program Excellence Initiative - Research University at the Jagiellonian University.

## Author contributions

All authors have accepted responsibility for the entire content of this manuscript and approved its submission.

## Competing interests

Authors state no conflict of interest.

## Informed consent

Informed consent was obtained from all individuals included in this study.

## Ethical approval

The local Institutional Review Board deemed the study exempt from review.